%% file: wisdom.tex
\documentclass{sig-alternate-2013}
\usepackage{multirow}
\usepackage{url}
\usepackage{array}

\usepackage{verbatim}
\usepackage{xcolor}

\newfont{\mycrnotice}{ptmr8t at 7pt}
\newfont{\myconfname}{ptmri8t at 7pt}
\let\crnotice\mycrnotice%
\let\confname\myconfname%

\permission{Permission to make digital or hard copies of all or part of this work for personal or classroom use is granted without fee provided that copies are not made or distributed for profit or commercial advantage and that copies bear this notice and the full citation on the first page. Copyrights for components of this work owned by others than the author(s) must be honored. Abstracting with credit is permitted. To copy otherwise, or republish, to post on servers or to redistribute to lists, requires prior specific permission and/or a fee. Request permissions from Permissions@acm.org.}
\conferenceinfo{HT'15,}{September 1--4, 2015, Guzelyurt, TRNC, Cyprus. \\
{\mycrnotice{Copyright is held by the owner/author(s). Publication rights licensed to ACM.}}}
\copyrightetc{ACM \the\acmcopyr}
\crdata{978-1-4503-3395-5/15/09\ ...\$15.00.\\
http://dx.doi.org/10.1145/2700171.2791056}

\begin{document}

\title{Wisdom of the Crowd or Wisdom of a Few?\\
An Analysis of Users' Content Generation}

\numberofauthors{2}
\author{
%
%
\alignauthor
Ricardo Baeza-Yates\\
	       \affaddr{Yahoo! Labs}\\
	       \affaddr{Barcelona, Spain}\\
       \email{rbaeza@acm.org}
\alignauthor
		Diego Saez-Trumper\\
	       \affaddr{Universitat Pompeu Fabra}\\
	       \affaddr{Barcelona, Spain}\\
	       \email{dsaez-trumper@acm.org}
}


\maketitle

\begin{abstract}
In this paper we analyze how user generated content (UGC) is created, challenging the well known {\it wisdom of crowds} concept. Although it is known that user activity in most settings follow a power law, that is, few people do a lot, while most do nothing, there are few studies that characterize well this activity. In our analysis of datasets from two different social networks, Facebook and Twitter, we find that a small percentage of active users and much less of all  users represent 50\% of the UGC. We also analyze the dynamic behavior of the generation of this content to find that the set of most active users is quite stable in time. Moreover, we study the social graph, finding that those active users are highly connected among them. This implies that most of the wisdom comes from a few users, challenging the independence assumption needed to have a wisdom of crowds. We also address the content that is never seen by any people, which we call digital desert, that challenges the assumption that the content of every person should be taken in account in a collective decision. We also compare our results with Wikipedia data and we address the quality of UGC content using an Amazon dataset. At the end our results are not surprising, as the Web is a reflection of our own society, where economical or political power also is in the hands of minorities.
\end{abstract}

;



\input{01_introduction.tex}
\input{02_related.tex}

\input{03_experimental_framework.tex}
\input{04_results.tex}

\input{05_quality.tex}

\input{07_conclusion.tex}

\newpage
\bibliographystyle{abbrv}
\bibliography{sigproc}

\end{document}

%% file: 01_introduction.tex
\section{Introduction}

The wisdom of crowds is a well known concept of how ``large groups of people are smarter than an elite few, no matter how brilliant, they are better at solving problems, fostering innovation, coming to wise decisions, and even predicting the future'' \cite{wisdom-of-crowds}.
On the other hand, although all people that use Internet can contribute to web content (or any type of activity), most people do not.
In fact, in any social network, the set of people that just looks at the activity of others (passive users or {\it digital voyeurs}) is much larger than the people that is active. Similarly, among the active users most of them do little, while a few do a lot ({\it digital exhibitionists}). We are interested in the characterization and interplay of these groups of people regarding the generation of content.

Let us take a specific case, say the world of blogs in the Web. Most people do not have a blog and few people have good blogs. Conversely, most blogs are not read and few blogs are well read. Indeed, people contribute to content in a social network or in the Web because they have the (possibly wrong) perception that someone will look at and read their contribution. This perception that they are speaking to the whole world, when the truth is that most of the time they are speaking alone, creates a very long tail of content that nobody sees, a huge {\it digital desert} where people write to an empty audience, metaphorically speaking.

Although we believe that there is a high correlation between the quality of content and the activity of users interacting with that content, in this paper we explore this process: how people contributes to content and what is the impact of the
content generation process in the so called wisdom of crowds. As we cannot study this in the context of the whole Web, as most usage data is private, we use two different datasets: a small sample of New Orleans Facebook users and a large one coming from a micro-blogging platform, Twitter. Both are good case studies for the problem being tackled. In fact, today Facebook and Twitter are the two largest social networks in term of users. In one of these cases we estimate a weak lower bound of how much of the UGC produced is never seen.  We also compare the content generation process in these social networks to the content generation of Wikipedia as well as the estimations of unique users per month that visit that website. 

\sloppy
Moreover, using another UGC dataset from Amazon's movies reviews, where the quality of content it is explicitly rated, we study the relation between quantity and quality of content produced by people, finding also that the majority of high-quality content is generated by a small set of users. 

Our main results are:
\begin{itemize}  \itemsep0em
\item The percentage of users that generate more than 50\% of the content is small, less than 7\% in our two examples;
\item These top users are quite stable in time, more than 70\% of the initial people in our two examples stay on that group during all the time observed;
\item The quality of content it is not strongly correlated with amount of users' activity, but;
\item Given that quality of content it is (almost) equality distributed among users, more active users produces - in absolute numbers - more high quality content than less active users.
\item The number of users that do not contribute to the generation of content is the majority of them, some because of inaction
while others because their content is not taken in account;
\item There is a significant volume of content that nobody sees, and hence is not taken in consideration; and
\item The bias seems to be even worse in non social contexts such as content creation in Wikipedia, where there are also is higher amount of content that is never visited.
\end{itemize}

The reminder of this paper is organized as follows. Sections 2 and 3 give the background. Sections 4 to 6 present the experimental results and discuss them. 


%% file: 02_related.tex
\section{Related Work}

The concept of the wisdom of crowds was introduced by Francis Galton in
1907~\cite{galton1907vox}, and used by James Surowiecki in his seminal book ``The
Wisdom of Crowds" \cite{wisdom-of-crowds}, where he posits --among other things-- that the aggregated knowledge of a group would be bigger than the knowledge of any of its single components. Although wisdom is difficult to measure, on the Web this concept has been translated --and widely applied-- as using the data provided directly ({\it e.g.}, content) or indirectly ({\it e.g.}, clicks) by users to discover knowledge in a \textit{crowd sourcing} approach~\cite{pacsca2007organizing,kittur2008harnessing,fuxman2008using,hsieh2013experts}. 
A good example of how this wisdom can be used, is exploiting the clicks that users do after issuing a query in a web search engine. This allows to extract semantic relations between queries in an automatic manner \cite{baeza2007extracting,Boldi:2008,cao2008context}. Therefore, in this example and others, more user generated content implies more knowledge that can be potentially discovered.

In Online Social Networks, wisdom can be related to the amount of content produced by users. Previous studies suggests that the amount of user's activity (\textit{e.g.}, number of \textit{tweets}) it is related with her/his number of followers~\cite{saez-trumper@ICWSM11}, and also with the monetary value that they produce~\cite{saez2014beyond}. Similarly, in social graphs --where node in-degree has a power-law distribution~\cite{Kwak,mislove-2007-socialnetworks,magno2012new,gonzalez2013google+}--  most of the content produced (\textit{i.e.}, activity) is generated by a small subset of users, while the majority of users act as passive information consumers~\cite{romero2011influence}. Moreover,  previous studies have shown that the around 50\% of URLs consumed in Twitter are produced by a tiny portion of users (less than 1\%) \cite{wu2011says}.  However, while previous work shows that to have a lot of followers cannot be considered as synonym of influence~\cite{falacy}, nowadays we do not know enough about most active users. In this paper we try to understand the importance and characteristics of most active users regarding the generation of content.

%% file: 03_experimental_framework.tex

\section{Experimental Framework}

\subsection{Assumptions and Definitions}
\label{sec:assumptions}
We consider that each unit of activity (tweets or posts) is {\em one unit of content} and that the overall activity is proportional to the wisdom of the crowd. A possible variation is to consider the length of the text of the tweet or the post. Nevertheless, as these texts are small (e.g. tweet length is capped by 140 characters), the results should be similar. Later on this paper, we discuss about the content's quality, and how it relates with the concept of wisdom. 

To distinguish top contributors ({\it wise users} or digital exhibitionists) from the rest of the active users, we use the following arbitrary definition for a given time period: wise users are the set of most-active users such that they contribute with 50\% of the content (or half of the wisdom). Other definitions are possible, for example based in a larger percentage, but we consider that 50\% is already a majority of the content. Nevertheless, the results would be similar as all the distributions involved resemble power laws. We call the rest of the users, in fact, the majority of them, {\it others}. 

\subsection{Datasets}

We use two different datasets from two different kind of social networks: Twitter, a micro-blogging social network; and Facebook, a pure social network. For all the experiments we consider only the active users, meaning users that have shown some posting activity in the time period considered.

\mbox{ } \\
\textbf{Facebook:}
 This  dataset corresponds to the New Orleans Facebook's Regional Network\footnote{Regional Networks were deprecated by Facebook in August of 2009.}~\cite{viswanath-2009-activity}.  We have two lists: the first one contains the social graph (friendships)  and a second list with user-to-user {\it wall posts} (where \textit{u,v} means user \textit{v} posting in \textit{u}'s wall), and the timestamp. All these data has been anonymized.

The social graph has 1,545,686 edges between 63,731 users.
The information about {\it wall posts} has 876,993 actions, with 39,986 users doing at least one post (active users). Hence, we can estimate that at least 37\% of users are passive or inactive. Notice that these are users that have a public profile, so although the set is partial, is what can be compared to other datasets based on public data as the next one.

This dataset has {\it wall posts} from 14th of September 2004 to 22th January of 2009 according to Table~\ref{tab:Facebook}.
We use the last three years as the two first are too small. The Pearson correlation of the number of posts with respect to the number of friends, using a logarithmic transformation to linearize the distributions, is 0.64. That is, the distributions are partially correlated.
The distribution of posts versus users follows a power law of parameter -1.58.

\begin{table}[!t]
\centering
\begin{tabular}{|l|c|c|}\hline
Time Period      & \#Active Users& \#Posts \\ \hline\hline
1 Year ~(2006)    &73K  &  8K         \\
2 Years (2006-2007)  &304K  &  18K        \\
3 Years  (2006-2008) &448K    &  18K        \\ \hline
All ~~~~~~(2004-2009)  &876K  &  39K       \\ \hline
\end{tabular}
\caption{Details of the Facebook dataset.}
\label{tab:Facebook}
\end{table}
\mbox{ } \\
\textbf{Twitter:} Our dataset contains almost all the tweets done in Twitter between March 1st until May 31st of 2009. We also have the complete social graph of Twitter for that period. This information is a subset of the dataset obtained in \cite{falacy}. Specifically, tweets are represented as a list of pairs (user id, timestamp), and the social graph is an adjacency list. 

The social graph has 1,963 millions of edges between 42 million of users (688 million edges between 12 millions of active users).
The activity considers 440 million of tweets produced by the active users. In fact, the number of non-active users (71\%) is more than 2.4 times larger than the number of active users (29\%). Hence, our analysis would be more striking if we take percentages over the whole user population. 

Our Twitter data has two limitations: (i) we have  ``only'' the last 3,200 tweets from each user, but we have found only 167 of users with more tweets than this threshold in around 50 millions users;\footnote{In any case, this implies that our results are a good lower bound because we are trimming the most active users.} and (ii) from the social graph, we cannot establish when each edge was created, therefore we are working with the final snapshot of that graph. 

In order to study the UCG with different time granularity, in our experiments we use this dataset in two different ways: first, the full dataset split in months and next, a smaller sample where we split in weeks the first three weeks of May. Tables~\ref{dataset} and \ref{dataset1} gives the details of them. The distribution of tweets versus users can be approximated by a power law of parameter -2.1. On the other hand, the Pearson correlation of the number of tweets with respect to the number of followers, using again the logarithmic transformation, is 0.68. That is, the distributions are again partially correlated.

\begin{table}[t!]
	\centering
\begin{tabular}{|l|c|c|c|}\hline
Time period  & \# Users & \# Active Users & \# Tweets \\ \hline\hline
Month 1 & 28M & 4.7M (16.7\%) & 109M \\
Month 2 & 37M & 7.3M (19.7\%) & 164M \\
Month 3 & 42M & 6.9M (16.4\%) & 167M \\ \hline
All     & 42M & 12.1M (28.8\%) & 440M \\ \hline
\end{tabular}
\caption{Details of the Twitter data set.}
\label{dataset}
\end{table}

\begin{table}[t!]
	\centering
\begin{tabular}{|l|c|c|c|}\hline
Time period  & \# Users & \# Active Users & \# Tweets \\ \hline\hline
1 Week  & 39M & 3.9M (10.0\%) & 48M \\
2 Weeks & 40M & 5.4M (13.5\%) & 95M \\
3 Weeks & 41M & 6.4M (15.6\%) & 141M \\ \hline
\end{tabular}
\caption{First three weeks of May, cumulative, for the small Twitter dataset.}
\label{dataset1}
\end{table}

\begin{figure}[tb!]
\centering
\epsfig{file=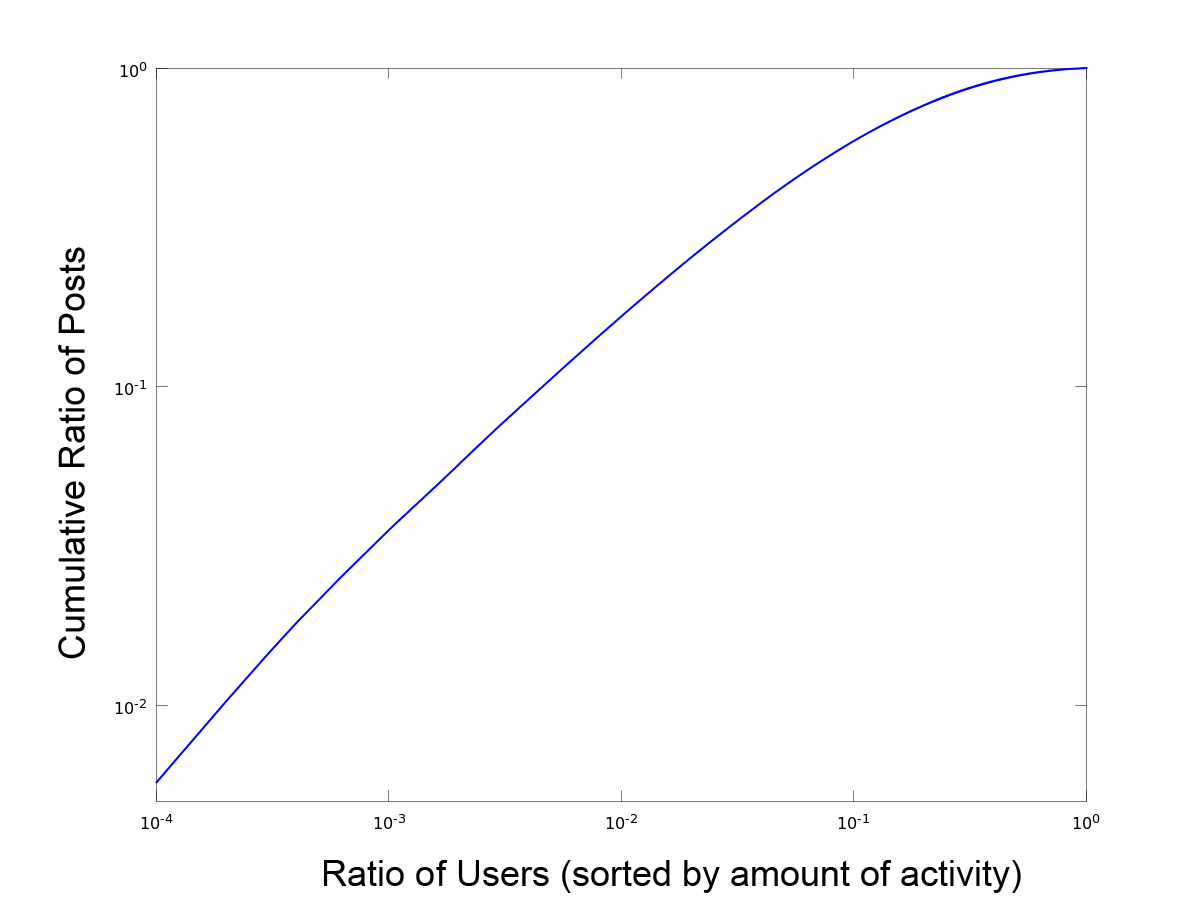, width=.49\linewidth}
\epsfig{file=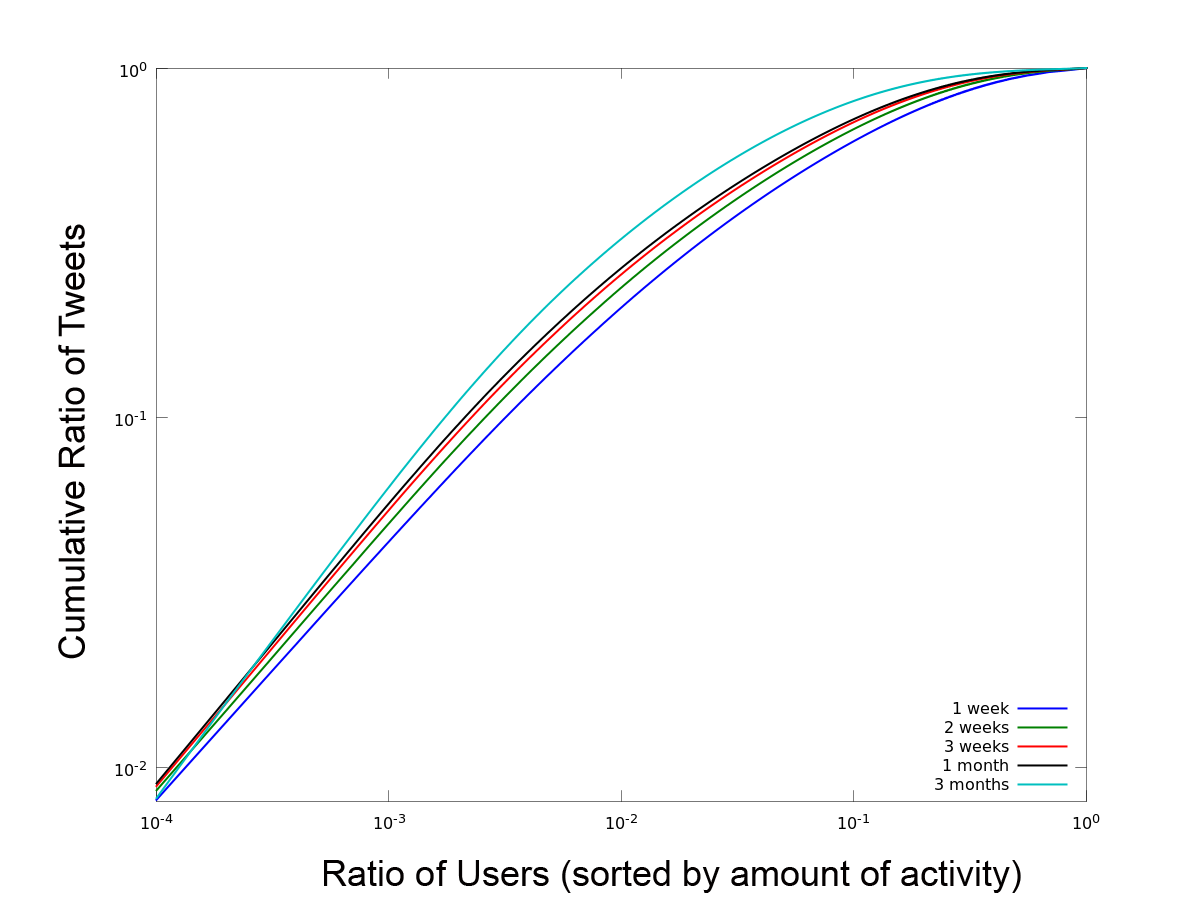, width=.49\linewidth}
\vspace*{-0.5cm}
\caption{Cumulative distribution of user activity in Facebook (left) and Twitter for the 3 weeks dataset (right).}
\label{fig:distro}
\vspace*{-0.3cm}
\end{figure}

%% file: 04_results.tex
\begin{table}[t!]
\centering
\begin{tabular}{|l|c|c|c|c|}\hline
          & \multicolumn{2}{c|}{Average in-degree} & \multicolumn{2}{c|}{Gamma index} \\ \cline{2-5}
Group     & Facebook & Twitter & Facebook & Twitter  \\ \hline\hline
Wise      & 100.5    & 1723.5 & $9.7\times10^{-3}$& $7.0\times 10^{-5}$ 	   \\
Others    &  29.1    &   81.5  & $6.2\times10^{-8}$& $2.4\times 10^{-6}$ \\ \hline
Active    & 34.1     & 112.6  & $3.8\times10^{-4}$& $4.6\times10^{-6}$\\ \hline
\end{tabular}
\caption{Average in-degree and Gamma index by group.}
\label{tab:avg_indegree}
\end{table}

\section{Experimental Results}

\subsection{Wise and Others}

We start by finding the proportional sizes of the user groups defined in the previous section. Figure \ref{fig:distro} shows that the distribution of user activity is very skewed. For Twitter - Figure \ref{fig:distro} (right) - where we have more data points, we show that the distribution also depends on the time window considered, as a longer time window implies that a smaller group of users produced most of the content. For three years, we found that in the Facebook dataset, the wise users were just 7.0\% of the total. On the other hand, in a period of three months, we found that in the Twitter dataset the wise users accounted for just 2.4\% of them. 

In the case of the Twitter dataset, looking at the social graph we found that even though wise users are less than 3\%, they concentrate more than the 35\% of the incoming edges, therefore they have also a higher in-degree (see Table \ref{tab:avg_indegree}). On the other hand, in the Facebook dataset, 7\% of the users concentrate 21\% of the links. This is not surprising as the Facebook social graph is more sparse as friendship is bidirectional and both users have to accept the relation.

To measure the connectivity of each group, we use the Gamma index, that is the ratio between the links observed over all possible links in  the complete graph of active users \cite{ricotta2000quantifying}. A larger Gamma Index means higher connectivity into the graph. In Twitter we see that  wise are the most cohesive group. In Facebook the differences are even bigger, as wise users are five orders of magnitude more cohesive than the rest. Differences between Twitter and Facebook might be due to the different nature of the link creation process in each platform (in Facebook both parts needs to agree to create a link, while in Twitter each user can decide alone) and the type of graph. However, in both cases wise users are more cohesive than the rest, suggesting that they are a  highly connected elite.


We can partially compare these results to the content generation process of Wikipedia. Indeed, according to data published by Wikipedia itself, the top 10,000 editors produce 33\% of the content editions. Considering that there are almost 20.8 million registered editors, the top editors represent just 0.04\% of them. 
As the number of passive users, that is, people that use Wikipedia but do not contribute with content, is more than one billion, the percentage of active editors with respect to the total number of users is negligible. Something similar happens with the creation of the almost 4.5 million Wikipedia articles in English, where the 2,005 most prolific authors account for the creation of 50\% of the articles~\cite{fewwikipedia}. This is less than 0.01\% of registered editors, and this number would be even smaller if non-registered users could be taken into account. 

\pagebreak
\subsection{Evolution Along Time}

Now we find the percentage of wise users for different periods of time for both datasets. Results are detailed in Tables \ref{tab:wise_per_year_facebook} and \ref{tab:wise_per_month}. This shows that even though the percentage of wise users decreases with larger time windows, the absolute number is pretty stable. However, are those wise users always the same? Table \ref{tab:wise_from_beginning_facebook} shows the percentage of users that were in the wise group in the first year and stay there in the next two years for the Facebook dataset.
Table \ref{tab:wise_from_beginning} shows the percentage of users that were in the wise group in the first week and stay there during the next two weeks for the small Twitter dataset. As can be seen the wise users are very stable, as more than 70\% or 80\% remains after three years or weeks, respectively.

\begin{table}[!t]
\centering
\begin{tabular}{|l|c|c|}\hline
Class      & Wise (\%) & Total Active Users \\ \hline\hline
1 year     &906 (11,8\%)   &  7663         \\ \hline
2 years     &1528 (8,7\%)   &  17572         \\ \hline
3 years     &2591(6,8\%)   &  38144        \\ \hline
\end{tabular}
\caption{Wise users per year (Facebook dataset).}
\label{tab:wise_per_year_facebook}
\end{table}
\begin{table}[t!]
\centering
\begin{tabular}{|l|c|c|}\hline
Time period     & Wise (\%) & Active Users \\ \hline\hline
1 week     & 231,620 (5,8\%)   &  3,979,668    \\ 
2 weeks   & 247,435   (4,7\%)&  5,362,828 	   \\ 
3 weeks   & 254,616   (4,0\%) &  6,415,867 	   \\ \hline
1 month   & 256,663 (3,7\%)   &  6,943,311	   \\ \hline
3 months   & 294,230 (2,4\%) &   12,183,943	   \\ \hline
\end{tabular}
\caption{Wise users in different time periods for the Twitter dataset.}
\label{tab:wise_per_month}
\end{table}

\begin{table}[!t]
\centering
\begin{tabular}{|l|c|c|}\hline
Class      & Wise from the beginning  & (\%) \\ \hline\hline
1 year     & 906 & 100       \\ \hline
2 years    & 671  &       74   \\ \hline
3 years    & 653  &       72   \\ \hline
\end{tabular}
\caption{Users that were in the wise group in the first year and stay there (Facebook). }
\label{tab:wise_from_beginning_facebook}
\end{table}
\begin{table}[t!]
\centering
\begin{tabular}{|l|c|c|}\hline
Time period         & Wise from the beginning  & (\%) \\ \hline\hline
1 week     & 231,620 & 100  	   \\ \hline
2 weeks   & 203,098  & 	87   \\ \hline
3 weeks   & 192,870  & 	83   \\ \hline
\end{tabular}
\caption{Users that were in the wise group in the first week and stay there (Twitter). }
\label{tab:wise_from_beginning}
\vspace*{-0.3cm}
\end{table}

\begin{figure*}[t!]
\centering
\epsfig{file=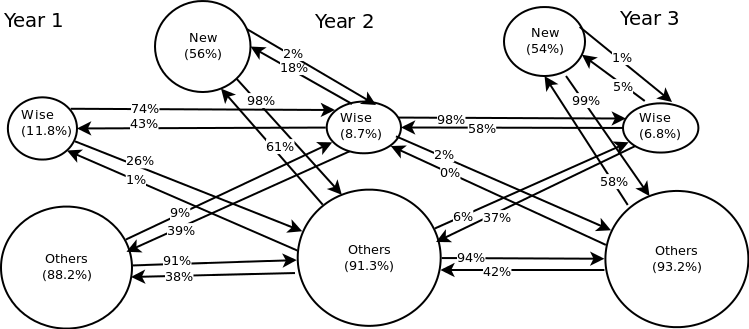, width=0.48\linewidth}
\centering
\epsfig{file=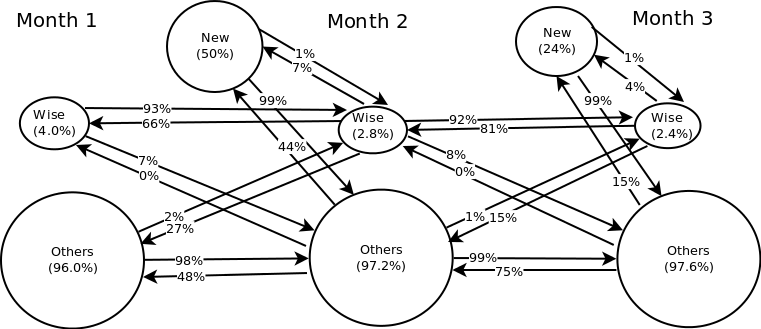, width=0.48\linewidth}
\vspace*{-0.2cm}
\caption{Dynamic behavior of users' groups for Facebook (left) and Twitter (right). Labels on edges represent the percentage of users coming/going from/to a given group to another.}
\label{fig:dynamics}
\end{figure*}

In Figure~\ref{fig:dynamics} we show the dynamics of the wise and others groups, during three years or months for both datasets, showing the percentage of people that come from the groups in the previous month as well as the percentage of new users. The numbers displayed in the edges, represents the percentage of users going to a given group in previous/next time slot. Outgoing edges pointing to previous time slot ({\em e.g.} from month 2 to month 1) shows where the users come from, while  edges pointing to the next time slot shows the destiny of those users.  For example, in Figure~\ref{fig:dynamics}~(b), in month 2, 66\%  of wise users come from the wise group in month 1, 27\% from  others and, 7\% from {\em new} users (users that were not active in month 1). Next, 92\% of those  wise users, stay in the  wise group  in month 3, and 8\% went to {\em others}. Another way to understand this, would be look at symmetric edges. For example, in month 3, 1\% of the {\em new} users went to wise group, while 4\% percent of the total of wise users come from {\em new} users. 

Overall, we can see that groups are quite stable if we do not consider new users. In fact, at the end of three periods, most of the {\em wise} users have always been in this group, for example in Figure~\ref{fig:dynamics}~(a), 74\% of {\em wise} users stay in that group from year 1 to year 2, and then 98\% stay in that group from year 2 to year 3, confirming the stability of this group.

\subsection{The Digital Desert}

In this section we analyze the phenomena of the content that is uploaded for some users, but is never seen by anyone else. We refer to this content as the {\em digital desert}. We can estimate a lower bound for the content that is never seen for the case of the Twitter dataset.\footnote{For the Facebook Dataset we cannot estimate the size of the digital desert because was obtained through a snowball sampling.} In fact, a lower bound for the digital desert can be computed as the percentage of content generated by people that has no followers.\footnote{Potentially, content uploaded by users without followers can be reached through the search page of Twitter or a generic search engine. However, tweets posted by users without followers are unlikely to be top-ranked in any search results.} This percentage of people is only 0.06\% for the wise group but 20.58\% for the {\it others} group. This accounts for 0.03\% and 1.08\% of the whole content, respectively. Hence, the digital dessert in this dataset is at least 1.11\%. 
Although small, this implies that the opinion of some people is not really considered and hence is not part of the collective wisdom. 

The size of the digital desert increases if we look at another kind of UGC platform such as Wikipedia. Comparing the logs of requested pages in the English Wikipedia  during a month (June 2014)\footnote{{https://dumps.wikimedia.org/other/pagecounts-raw/}} with the new content added in the previous month (May 2014) we see that from the 1,350,554 articles edited/added during that month, 31\% of them were not visited\footnote{These visits include humans and bots.} in June. This is an upper bound for the digital desert in this dataset and time period.

%% file: 05_quality.tex
\section{Quality of the Wisdom}
\label{sec:quality}

In previous sections we have used an arbitrary definition of wisdom that is directly related with the amount of content produced by users. However, one can argue that quantity of content  produced ({\em i.e.,} activity) does not imply equal contribution to the global wisdom. To address this problem we need to measure the quality of content. Unfortunately, it is not simply to measure content quality in a social network, because it would be difficult to define what is a ``good tweet" or a ``good post" in Facebook. One option is to relate quality with popularity ({\em e.g.}, retweets or likes), but such metric would be clearly biased towards popular users. Therefore, it is preferable to use a dataset where the quality of users' contributions it is clearly ranked by the readers. A good example of such kind of content are Amazon's products reviews, where readers can evaluate the helpfulness of a review by answering yes or no to the following question: ``Was this review helpful to you?"

\sloppy
Specifically, we use a public Amazon's movie reviews dataset released by \cite{mcauley2013amateurs} in 2013. This dataset contains almost 8 million reviews, from 889,176 users, of around 250K different movies, in a period of 15 years (from 1997 to 2012).  From each review we have --among other things-- the (anonymized) author, the content, and also a field called ``helpfulness", that contains the number of readers that have rated the review as helpful or not. 

In order to make this data comparable with the previous experiments, first we divided users in wise and others, following the definition given in Section \ref{sec:assumptions} that based in the amount of activity (previously number of post or tweets, now number of reviews). In this case we found that 4\%  of users produced 50\% of all reviews. This is similar to Facebook (7\%) and Twitter (2\%), suggesting that the process of content generation is comparable with the previous cases. For future comparison we denote this group of users as activity-based-wise. 

Next, we want to redefine wisdom by adding the dimension of content's quality. To do that, we say that users are contributing to the  wisdom only if each review has been rated as helpful by at least one reader. The intuition behind this definition is that if a review helped at least one user, the review is a contribution to the total wisdom. Obviously, stronger requirements can be imposed ({\em e.g.}, that at least 50\% of the users rating a review found it useful). However, our definition will establish a lower bound for the content's value. Hence, now the total wisdom will be the sum of all helpful reviews. Surprisingly, we found that 64\% of the reviews were helpful for at least one reader, and  66\% of users have produced at least one helpful review, showing that a wide group of users contribute to the total good content and that almost two thirds of the whole wisdom generated is valuable. However, breaking down the results we found that --again-- just 2.5\% of the users produced 50\% of the total helpful reviews. We denote these users as quality-based-wise. Moreover, we found that quality-based-wise users is a {\em proper} subset of the activity-based-wise users in this dataset.


\begin{figure}[t!]
\centering
\epsfig{file=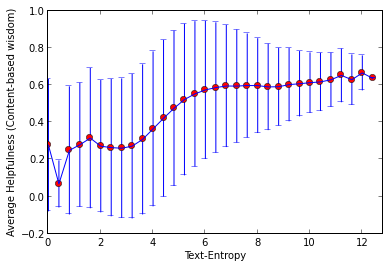, width=0.75\linewidth}
\vspace*{-.3cm}
\caption{Average Helpfulness of users according to text entropy in their reviews.}
\label{fig:entropy-helpfulnes.png}
\vspace*{-.3cm}
\end{figure}



We also compute the (review) entropy for each user. To that aim, we grouped the reviews of each single user, and then computed the Shannon entropy in that text. Interestingly, we found that the Spearman correlation between activity and entropy is low (0.32), while the correlation between entropy and  helpfulness is slightly higher (0.43). Figure~\ref{fig:entropy-helpfulnes.png} shows that from a certain level of users' entropy the reviews tend to be more useful, but that also there is a saturation point where more entropy does not imply more helpfulness. This relation between entropy and value (helpfulness) is useful to generalize  these results because we expect that users that introduce more information per word ({\em i.e.}, higher text entropy) are --at the same time--  adding more wisdom.

%% file: 07_conclusion.tex
\vspace*{0.15cm}
\section{Conclusions and Future Work}

Our results, added to social influences, undermines the independence principle that is needed to have a real wisdom of crowds \cite{wisdom-of-crowds}, as the percentage of people that produces most of the content is really small. Moreover, if we consider that very active people is highly connected among them, compared with the rest of users, creating a cohesive elite. The diversity principle is also challenged, as many users do not contribute to the wisdom, either because they do not exercise this option or because their opinion is not taken in account (the digital desert).

The distribution of how people contribute to wisdom becomes more skewed when a filter of quality is introduced. Although  many people show the capability of producing helpful content, the majority of such content is produced by just a subset of the elite. 

The datasets used are already a bit old, but on the other hand one of them is complete and less noisy than a more current dataset, as at that time Twitter had less spam than nowadays. For sure the results in other datasets would be different, but we believe the issues addressed in this paper will remain valid for the majority of UCG. 


Finally, although we would like to believe that the Web is a more democratic environment as all the people has the same opportunities,
at the end the Web mimics our society. Indeed, the economic or political power in most countries belongs to a minority of the people.
Even when explicit decisions must be taken through elections or referendums, many people choose not to exercise their right to vote.